\documentclass[aps,twocolumn,showpacs]{revtex4}
\usepackage{amsmath}
\usepackage{graphicx}
\usepackage{amsfonts}

\begin{document}

\bibliographystyle{apsrev}

\newcommand{\ket}[1]{\left| #1 \right>}
\newcommand{\bra}[1]{\left< #1 \right|}

\title{Modified Grover's algorithm for an expectation value quantum computer}
\author{David Collins}
\affiliation{Department of Physics, Carnegie Mellon University, 
             Pittsburgh PA 15213 }
\email{collins5@andrew.cmu.edu}

\date{\today}

\begin{abstract}

 The translation of Grover's search algorithm from its standard
 version, designed for implementation on a single quantum system
 amenable to projective measurements, into one suitable for an
 ensemble of quantum computers, whose outputs are expectation values
 of observables, is described in detail. A filtering scheme, which
 effectively determines expectation values on a limited portion of the
 quantum state, is presented and used to locate a single item for
 searches involving more than one marked item. A truncated version of
 Grover's algorithm, requiring fewer steps than the translated
 standard version but locating marked items just as successfully, is
 proposed. For quantum computational devices which only return
 expectation values, the truncated version is superior to its standard
 counterpart. This indicates that it is possible to modify quantum
 algorithms so as to reduce the required temporal resources by using
 the ensemble's spatial resources.
\end{abstract}

\maketitle

\section{Introduction}

The standard formulation of quantum algorithms, such as the
Deutsch-Jozsa \cite{cleve}, Shor factorization \cite{shor} and Grover
search algorithms \cite{grov}, assumes that the algorithm will
eventually be implemented on a single quantum system amenable to
projective measurements. Currently, however, the most advanced
realizations of quantum algorithms use room temperature solution state
NMR, where an entire ensemble of quantum systems must be manipulated
and which can only yield expectation values of certain observables
\cite{CFH,ChGKL,ChGK98,JMH98,linden98,VYSCh99,VSSYBCh00,marx00,cummins00,
collins00,kim00,vandersypen00,weinstein01,long01}. Consequently
various steps of the standard algorithm formulation have to be
modified for implementation on such an expectation value quantum
computing device. The typical approach has been to translate the
operations as literally as possible so that the expectation value
quantum computer effectively mimics its single quantum system
relative. In this article we consider this process for Grover's
algorithm and report a superior and significantly different version,
which effectively uses the ensemble's spatial resources to reduce the
required temporal resources, for implementation on an expectation
value quantum computer.

We conclude this introduction with a description of a general scheme
for translating the standard version of a quantum algorithm into a
form suitable for realization on an expectation value quantum
computer. In section \ref{sec:grover} we briefly describe the standard
formulation of Grover's algorithm. The main result of this article is
contained in section~\ref{sec:oneitem} which first describes a
standard expectation value version of Grover's algorithm for searching
a database containing one marked item and then proposes a truncated
version requiring fewer steps. Section~\ref{sec:more_than_one} extends
these ideas to situations where there is more than one marked item and
proposes a scheme of filtered expectation values for extracting the
location of a single marked item. Finally in section~\ref{sec:other}
we briefly consider initial attempts to apply these techniques to
other quantum algorithms.

In the standard formulation of any quantum algorithm the first step is
to prepare a single $L$ qubit quantum system in a pure state
$\ket{\psi_i}$. This is followed by applying of a sequence of unitary
time evolution transformations, $\hat{U}_1, \ldots, \hat{U}_m$,
performing $\ket{\psi_i} \rightarrow \ket{\psi_f} := \hat{U}_m \ldots
\hat{U}_2\hat{U}_1 \ket{\psi_i}$. The time evolution transformations
collectively form a single \textit{algorithm unitary operator},
$\hat{U}_{\text{alg}}:= \hat{U}_m \ldots \hat{U}_2\hat{U}_1$. Finally
a computational basis projective measurement (PM) is performed on an
$L'$ qubit subset of the quantum system resulting in $x \in
\{0,\ldots,2^{L'}-1 \}$. The corresponding collection of measurement
operators are the projectors $\{ \hat{P}_x = \ket{x} \bra{x} \; | \;
x=0,\ldots,2^{L'}-1 \}$ and the output $x$ is produced with
probability $\bra{\psi_f} \hat{P}_x \ket{\psi_f} = \left| \left<x |
\psi_f\right>\right|^2$. Typically this is nonzero for more than one
value of $x$, some of which may yield incorrect solutions to the
problem at hand. However, efficient quantum algorithms have the
property that the probabilities of such failures are appropriately
small; Shor's algorithm offers an example \cite{shor}.

In expectation value quantum computation, such as standard NMR
implementations, there are two important differences. First, the
single quantum system is replaced by an ensemble and in the case of
solution state, room temperature NMR this is in a highly mixed
state. However, various pseudo-pure state preparation schemes
\cite{ChGKL,GC,KChL97,cory98} result in a state that, as far as
typical NMR measurements are concerned, resembles a pure state
although it may lack the entanglement properties associated with the
pure state \cite{schack99}. Such preparation schemes effectively carry
out the same initialization required by the single, pure state quantum
system formulation. Second, the most basic outputs which an
expectation value quantum computer can provide are expectation values
of observables over the entire ensemble; this is the case with NMR
quantum computers. This stands in contrast to situations where
individual quantum systems are accessible, in which case the basic
outputs are the results of projective measurements which can then be
used to derive expectation values. A translation between the PM and
expectation value (EV) protocols is essential for the development of
expectation value quantum computation. This is accomplished by
considering PMs on an ensemble of identically prepared quantum
systems.  Firstly a PM can be accomplished in a bitwise fashion via
projective measurement in the basis $\{ \ket{0}, \ket{1} \}$ for each
qubit. The resulting outcomes yield the expectation values of
$\hat{P}^{(k)}_0 =\ket{0}\bra{0}$, from which the expectation value of
$\sigma_z^{(k)} = 2 \hat{P}^{(k)}_0 - \hat{I}^{(k)}$ is easily
computed (here $k$ labels the qubit). The comparative ease with which
Pauli operators are manipulated versus computational basis projection
operators results in the standard practice of replacing the PM by the
EVs of the set of single qubit observables $\{ \sigma_z^{(k)} \; | \;
k=1,\ldots, n \}$. Henceforth we assume that it is possible to measure
the expectation value of $\sigma_z^{(k)}$ for an arbitrary
qubit. Finally we note that it is generally assumed that the algorithm
unitary operator (or at least its constituents) can be translated
straightforwardly into a sequence of operations suitable for an
expectation value quantum computer. This is certainly the case for
standard NMR quantum computers \cite{CFH,ChGKL,cory98}

The two protocols are equally powerful for reversible deterministic
classical computation, in which case $\ket{\psi_f} = \ket{s}$ for some
$s \in \{ 0, \ldots, 2^n-1\}$. The standard EV yields
\begin{equation}
 \left< \sigma_z^{(k)} \right> \equiv \bra{\psi_f} \sigma_z^{(k)} 
                                      \ket{\psi_f}= (-1)^{s_k}
 \label{eq:classical_output}
\end{equation}
where $s= s_n\ldots s_2s_1$ is the bitwise representation of the
algorithm solution. On the other hand, the PM extracts $s= s_n\ldots
s_2s_1$ directly and the equivalence is clear.  The only requirement
is the ability to distinguish between $\left< \sigma_z^{(k)} \right> =
+1$ and $\left< \sigma_z^{(k)} \right> = -1$ for each qubit. In
quantum computation $\ket{\psi_f}$ is often a superposition of
computational basis states, each of which contribute to $\bra{\psi_f}
\sigma_z^{(k)} \ket{\psi_f}$. In some cases, such as those presented
in section~\ref{sec:more_than_one} or ref~\cite{GC}, these
contributions cancel and inferences such as those offered in
Eq.~(\ref{eq:classical_output}) no longer apply. It is not surprising
that the single quantum system, pure state versions of algorithms may
have to be modified so as to be amenable to implementation via
expectation value quantum computers. Indeed, this modification has
already been provided for Shor's algorithm \cite{GC}.

\section{Grover's search algorithm}
\label{sec:grover}

Grover's quantum algorithm for searching an unstructured database
provides a quadratic speedup compared to the classical sequential
search \cite{grov,NC}. The relatively small computational resources
required (compared to Shor's algorithm) has motivated several
successful small scale experimental implementations, using NMR quantum
computers, for databases which conceal one marked item in one of as
many as eight locations
\cite{ChGK98,JMH98,VYSCh99,VSSYBCh00,cummins00,long01}. The primary
focus in these demonstrations appears to have been the successful
implementation of the preparation and algorithm unitary operator
steps; here the typical measurement operations aimed to verify the
accuracy of the aforementioned preceeding steps.

As usual the standard version of the algorithm is formulated in terms
suitable for single quantum systems amenable to projective
measurements. The problem involves a database which conceals $M$
marked items in $N$ possible locations, where it is assumed that $M$
is known (note that there exists a quantum algorithm, which is more
efficient than its classical counterparts, for finding $M$ if it is
not known in advance \cite{brassard98}). The task is to find the
location of any one of the marked items, given an oracle which can
answer the question, ``Is a marked item located at $x$?'' for any
possible location denoted $x$. The database will be represented as $X=
\{ 0,1, \ldots , N-1 \}$ and the locations of the marked items form a
subset, $S \subset X$. Querying the oracle is equivalent to evaluating
\begin{equation} 
f(x) :=
 \begin{cases}
  0 & \text{if $x \notin S$} \\
  1 & \text{if $x \in S$}
 \end{cases}
\end{equation}
and it is assumed that this can be done at unit cost for any $x \in
X.$ A classical sequential search evaluates $f(x)$ for a succession of
distinct $x \in X$ until it returns $1$. For $M \ll N$ this requires
approximately $N/(M+1)$ oracle queries on average to be certain of
locating at least one of the marked items.

The standard formulation of Grover's algorithm \cite{grov,NC}
represents database locations in terms of the computational basis
states, $\{ \ket{x}, x=0,1,\ldots , 2^L-1\}$, of an $L$ qubit data
register, where $L-1 \leq \log_2(N) \leq L$. Information about the
location of the marked items is supplied via an oracle unitary
operation, defined on the computational basis states as
\begin{equation}
 \hat{U}_f \ket{x} := (-1)^{f(x)} \ket{x}
\end{equation}
and extended linearly to superpositions of these. The database can be
extended to one with $2^L$ locations by requiring that $f(x):=0$ for
$x = N,\ldots ,2^L-1$; henceforth we assume that $N=2^L$.  The data
register is initially prepared in an unbiased superposition of all
possible locations, i.e.\ $\ket{\psi_i} :=
1/\sqrt{2^L}\sum_{x=0}^{2^L-1} \ket{x}$, which is an unentangled state
and typically easy to create. Repeated applications of the Grover
iterate $\hat{G} := \hat{D} \hat{U}_f$ follow the preparation; the
crucial oracle unitary operations are interspersed with the unitary
``inversion about the average''
\begin{equation}
 \hat{D} \left( \sum_{x=0}^{N-1} c_x \ket{x} \right) 
   := \sum_{x=0}^{N-1} \left( -c_x + 2 \left< c \right> \right) \ket{x}
\end{equation}
where $\left< c \right> =\sum_{x=0}^{N-1} c_x / N$.  Analysis of the
algorithm is simplified by noting \cite{bbht98,zalka99} that, after
any number of applications of $\hat{G}$, the state of the data
register has the form
\begin{equation}
\ket{\psi} := \frac{\alpha}{\sqrt{M}} \sum_{x \in S} \ket{x} + 
              \frac{\beta}{\sqrt{N-M}} \sum_{x \notin S} \ket{x}
\label{eq:gen_grover_state}
\end{equation}
where $\alpha,\beta \in \mathbb{R}$ satisfy $\alpha^2 + \beta^2 =
1$. Thus the data register state is conveniently represented by a two
dimensional real unit vector $\mathbf{v} := (\alpha,
\beta)$. Accordingly the probability with which a data register PM
will locate a marked item is $\alpha^2$. After $m$ applications of
$\hat{G}$,
\begin{equation}
 \alpha  =   \sin{\left[ (2m+1)\theta/2\right]} 
    \quad \text{and} \quad \beta = \cos{\left[ (2m+1)\theta/2\right] }
\end{equation}
where $\cos{\theta} = 1-2M/N$, and therefore each invocation of the
Grover iterate effectively rotates $\mathbf{v}$ through angle
$\theta$. The algorithm may be regarded \cite{zalka99,bbht98} as a
procedure for rotating the undesirable initial state $\mathbf{v}_i =
(\sqrt{M/N}, \sqrt{1-M/N})$ as close as possible to the desired final
state $\mathbf{v}_f = (1,0)$. If $N/M \gg 1$ then $\theta \approx
2\sqrt{M/N}$ and thus after $\lfloor \sqrt{N/M}\; \pi/4 \rfloor$
applications of the Grover iterate $\mathbf{v} \approx (1,0)$,
whereupon a PM will yield a marked item with probability close to
$1$. Rounding errors occur when there is no integer $m$ such that
$(2m~+~1)\theta/2 =\pi/2$ but even in the worst of these cases the
probability of success is at least $1-M/N$ \cite{bbht98}. Each
application of $\hat{G}$ queries the oracle once. Thus Grover's
algorithm provides a quadratic speedup compared to the classical
sequential search. We shall refer to this rendition of the algorithm
as the standard PM version.

\section{EV versions of Grover's algorithm: one marked item}
\label{sec:oneitem}

The first attempt to translate Grover's algorithm to a form suitable
for implementation on expectation value quantum computers results in a
standard EV version which essentially differs from its standard PM
counterpart only in the measurement stage. The translation is
considerably simpler whenever there is only one marked item than for
the remaining cases (i.~e.\ $M>1$), and the key ideas behind
modifications to the standard EV version are illustrated more clearly
in the former. Suppose that there is only one marked item, located at
$s$. In the standard EV version the ensemble is initialized via a
preparation scheme which results in a pseudo-pure state corresponding
to $\ket{\psi_i}= 1/\sqrt{2^L}\sum_{x=0}^{2L-1}\ket{x}$. The unitary
algorithm operator of the standard PM version is retained; only the
measurement stage must be altered. In the standard PM version the data
register is approximately in the state $\ket{\psi_f} = \ket{s}$ after
$\lfloor \pi \sqrt{N}/4 \rfloor$ applications of $\hat{G}$. The
algorithm is essentially deterministic and EVs of the single qubit
observables, $\sigma_z^{(k)}$ for $k =1, \ldots, L$, yield a bitwise
representation of $s$ as described earlier. Rounding errors result in
discrepancies with relative size at most $1/N$. Thus if there is only
one marked item the translation is accomplished by merely replacing
the projective measurements of the standard PM version with
expectation values of $\sigma_z^{(k)}$ and determining the binary
representation of the marked item's location by the sign of $\left<
\sigma_z^{(k)} \right>$ for each data register qubit. This is the
\textit{standard EV version} of Grover's algorithm. To date all
experimental NMR demonstrations of Grover's algorithm have taken this
approach \cite{ChGK98,JMH98,VYSCh99,VSSYBCh00,cummins00,long01}.

The standard EV version of the algorithm does not use the ensemble's
spatial resources (i.e.\ the replication of the quantum computing
device via the ensemble members) in any essential way. However, by
considering the effect of terminating the algorithm after
significantly fewer than $\lfloor \pi \sqrt{N}/4 \rfloor$ applications
of $\hat{G}$, we shall arrive at a \textit{truncated EV version} which
effectively reduces the required temporal resources by taking
advantage of these spatial resources and which still locates the
marked item with certainty.  It may be surprising that it could be
sensible to stop after significantly fewer than $\lfloor \pi
\sqrt{N}/4 \rfloor$ applications of $\hat{G}$ since this would leave
the data register in a superposition of computational basis states
somewhat distant from the ideal, namely $\ket{\psi_f} = \ket{s}$. At
the outset it is unclear whether, for such superpositions, the sign of
$\left< \sigma_z^{(k)} \right>$ could provide any useful information
regarding the location of the marked item. However, as we shall
demonstrate below, this is indeed a reasonable strategy. In fact,
whenever measurements only yield expectation values the resulting
approach will usually be superior to that of the standard EV version.

As with the standard EV version the system must be initialized in the
pseudo-pure state corresponding to $\ket{\psi_i}=
1/\sqrt{2^L}\sum_{x=0}^{2L-1}\ket{x}$. Then suppose that $\hat{G}$ is
repeatedly applied an \textit{arbitrary number of times}, resulting in
the state of the form given by Eq~(\ref{eq:gen_grover_state}). At this
stage data register EVs give (the discussion is presented for an
arbitrary number of marked items for later generalization)
\begin{equation}
 \left< \sigma_z^{(k)} \right>  =  \frac{ \alpha^2}{M} 
                                    \sum_{x \in S} (-1)^{x_k} +
                                   \frac{ \beta^2}{N-M} 
                                    \sum_{x \notin S} (-1)^{x_k}.
\end{equation}
However, 
\begin{equation}
 \sum_{x \in X} (-1)^{x_k} = 0 \quad \Rightarrow \quad 
                 \sum_{x \notin S} (-1)^{x_k} = - \sum_{x \in S} (-1)^{x_k}
\end{equation}
and thus
\begin{equation}
 \left< \sigma_z^{(k)} \right>  =  \left( \frac{\alpha^2}{M} 
                                 - \frac{\beta^2}{N-M} \right) 
                                   \sum_{x \in S} (-1)^{x_k}.
\end{equation}
Finally the normalization requirement, $\alpha^2 + \beta^2 =1$, yields
\begin{equation}
 \left< \sigma_z^{(k)} \right>  =  \frac{A}{M} \sum_{x \in S} (-1)^{x_k}.
 \label{eq:trunc_grover_state}
\end{equation}
where the \textit{EV attenuation}
\begin{equation}
 A : = \frac{\alpha^2N - M}{N-M}
\end{equation}
determines the magnitude of the EV. When there is only one marked item
and $\hat{G}$ has been applied $m$ times
\begin{equation}
 \left< \sigma_z^{(k)} \right>  =  \frac{A_m}{M}(-1)^{s_k}.
 \label{eq:trunc_grover_state_one}
\end{equation}
where the EV attenuation is 
\begin{equation}
  A_m = \frac{\sin^2{[(2m+1)\theta/2]}\; N - 1}{N-1}.
 \label{eq:amp_after_m}
\end{equation}
The binary representation of the marked item's location can still be
extracted by the usual process of inspecting the sign of $\left<
\sigma_z^{(k)}\right>$ for each data register qubit provided that
$A_m$ is sufficiently large. The latter increases monotonically with
respect to $m$ from $0$ (for $m=0$) to approximately $1$ in the case
of the standard EV version (for $m= \lfloor \frac{\pi}{4} \sqrt{N}
\rfloor$). Thus whenever the basic outputs are EVs it is entirely
feasible to modify the algorithm unitary operator so that it uses
fewer than $\lfloor\frac{\pi}{4} \sqrt{N} \rfloor$ applications of
$\hat{G}$ without reducing the probability of successfully locating a
marked item. The only effect of this truncation is to reduce the
magnitudes of the EVs by a factor of $A_m$ where $m$ is the number of
applications of $\hat{G}$.  Therefore we propose a \textit{truncated
EV version} of the algorithm which \textit{uses an $L$ qubit data
register initialized in the pseudo-pure state corresponding to
$\ket{\psi_i}= 1/\sqrt{2^L}\sum_{x=0}^{2L-1}\ket{x}$ and which
terminates after the minimum number of applications of $\hat{G}$ such
that it is still possible to distinguish reliably between $\left<
\sigma_z^{(k)} \right> >0$ and $\left< \sigma_z^{(k)} \right> < 0$ for
all data register qubits}. When there is only one marked item the
minimum number of applications of $\hat{G}$, $m$, is based on the
ability to distinguish between $\left< \sigma_z^{(k)} \right> = +A_m$
and $\left< \sigma_z^{(k)} \right> = -A_m$ for all qubits. The
resolution in measuring $\left< \sigma_z^{(k)} \right>$, and hence the
minimum tolerable number of applications of $\hat{G}$, usually depends
on features of the experimental setup such as the ensemble size and
the signal-to-noise ratio of the observed signal. However, in many
instances it should be possible to use significantly fewer invocations
of the oracle than required by the standard EV version without
reducing the probability of locating the marked item. In such cases
the truncated EV version is clearly superior to its standard EV
predecessor.

Proposals for expectation value quantum computing devices typically
assume that each of the required operations will be applied
simultaneously to all members of the ensemble; this is true for
existing NMR realizations. The truncated EV version of Grover's
algorithm reduces the required number of constituent operations of the
unitary algorithm operator, and hence the required temporal resources,
by effectively exploiting the ensemble's spatial resources. In this
sense too the truncated EV version represents a greater departure from
the standard PM version than the standard EV version does; not only
are the measurement stages but also the algorithm unitary operators
different.

The truncated EV version may appear to violate the optimality
\cite{bbbv97,bbht98,zalka99} of the standard formulation of Grover's
algorithm. However, it will only work if reliable expectation values
can be produced. This requires an ensemble of quantum systems and it
can be argued that each oracle invocation, via implementation of
$\hat{G}$, on the ensemble effectively amounts to one oracle query per
ensemble member. The truncated EV version of Grover's algorithm is not
intended to substitute the standard PM version. Rather in situations,
such as NMR, where the basic measurement quantities are expectation
values (in contrast to PMs) it is intended as an improvement on the
standard EV version.

\section{EV versions of Grover's algorithm: more than one marked item}
\label{sec:more_than_one}

Translating the standard PM version of the algorithm to EV versions is
significantly more complicated if there is more than one marked
item. Here, after $\lfloor \sqrt{N/M}\; \pi/4 \rfloor$ applications of
$\hat{G}$, the data register is approximately in the state
$\ket{\psi_f} = 1/\sqrt{M} \sum_{x \in S} \ket{x}$ (we henceforth
ignore the deviation from this approximation, which gives a
probability of $M/N$ of incorrectly locating a marked item, since
$M\ll N$). A PM will yield a location of one of the marked items with
certainty. However, merely substituting EVs for PMs, as was done for
the single marked item case, will sometimes not provide enough useful
information. An extreme example is that where there are two marked
items, whose locations have binary expansions, $s=s_L \ldots s_1$ and
$s'=s'_L \ldots s'_1$ where $s'_k = 1-s_k$ for all $k$. In this case
$\left< \sigma_z^{(k)} \right> = 0$ for all $k$. If it is known that
$M=2$ then the only information that this set of EVs provides is that
$s'_k = 1-s_k$. While it is possible to determine any single bit for
some $s \in S$, it is impossible to infer two or more bits of any $s
\in S$ with certainty. In general the strategy of replacing PMs with
EVs only offers the following limited information: if $\left<
\sigma_z^{(k)} \right> \geq 0$ then $s_k=0$ for some $s \in S$, while
$\left< \sigma_z^{(k)} \right> \leq 0$ implies that $s_k=1$ for some
$s \in S$.  However, such bitwise EVs ignore crucial inter-qubit
correlations, which are effectively exploited in the PM scheme to
yield the location of some marked item. The EV outputs for qubits $j$
and $k$ only give $s_j$ and $s'_k$ for some $s,s' \in S$. However, as
the example above illustrates, there does not necessarily exist any
single $s'' \in S$ whose relevant bits have these values. (i.~e.\ such
that $s''_j = s_j$ and $s''_k = s'_k$).

Thus we first develop an EV scheme, which exploits correlations
between qubits, to circumvent this obstacle and then use this to
provide a truncated EV version for Grover's algorithm when there is
more than one marked item.  The key idea, which will eventually allow
for a bitwise determination of a marked item's location using EVs, is
a \textit{filtered expectation value}, i.e.\ an EV on one qubit which
is conditional on the states of one or more of the remaining qubits. A
related technique has been proposed in connection with a logical
labeling scheme for pseudo-pure state preparation
\cite{ChGKL}. However, to the best of our knowledge, application of
this to situations such as encountered here has not been described
explicitly in the literature and the version offered here is somewhat
simpler to that of~\cite{ChGKL}.

To illustrate the filtered EV technique consider an instance where
$\left< \sigma_z^{(1)} \right> \leq 0$.  Then $s_1=1$ for some $s \in
S$. The intention is to determine $s_2$ for some $s \in S$ such that
$s_1=1$. The idea now is to repeat the algorithm so that, in the
course of measuring $\left< \sigma_z^{(2)} \right>$, the final data
register state is effectively filtered or restricted to $\ket{\psi_f}
= 1/\sqrt{M} \sum_{x \in S, x_1=1} \ket{x}$. Then the sign of $\left<
\sigma_z^{(2)} \right>$ gives $s_2$, and hence $s_2s_1$, for some $s
\in S$ such that $s_1=1$. The restriction could be accomplished by
redefining the oracle so as to fix $x_1=1$. However, this can be
avoided by comparing the outputs of two runs of the algorithm, the
second of which includes an additional unitary operation. First note
that after applying $\hat{U}_{\text{alg}}$, $\ket{\psi_f} = 1/\sqrt{M}
\sum_{x \in S} \ket{x}$ and
\begin{equation}
 \left< \sigma_z^{(2)} \right> = \frac{1}{M} 
                                 \sum_{x \in S, x_1=0}(-1)^{x_2} +  
                                 \frac{1}{M} 
                                 \sum_{x \in S, x_1=1}(-1)^{x_2}
 \label{eq:before_corr21}
\end{equation} 
Now suppose that $\hat{U}_{\text{alg}}$ is followed by a
\textit{correlation operation}, defined on computational basis states
and extended linearly to superpositions,
\begin{equation}
 \hat{C}_2(s_1=1) \ket{x_L \ldots x_1} 
      := \left(\sigma^{(2)}_x \right)^{x_1 \oplus 1}\ket{x_L \ldots x_1}
\end{equation}
where $x_k \in \{0,1\}$, and addition is modulo $2$. The argument of
$\hat{C}$ describes the filtering conditions and the subscript the
qubit whose EV will be determined subject to such conditions. Applying
this after the unitary algorithm operation gives
\begin{equation}
 \ket{\psi_f} \stackrel{\hat{C}_2(s_1=1)}{\longrightarrow} 1/\sqrt{M} 
                              \sum_{x \in S} 
                              \left( \sigma_x^{(2)} \right)^{x_1\oplus 1} 
                              \ket{x} 
\end{equation}
after which
\begin{equation}
 \left< \sigma_z^{(2)} \right> = - \frac{1}{M} 
                                   \sum_{x \in S, x_1=0}(-1)^{x_2} 
                                 + \frac{1}{M} 
                                   \sum_{x \in S, x_1=1}(-1)^{x_2}
 \label{eq:after_corr21}
\end{equation} 
since $\sigma_x^{(2)} \sigma_z^{(2)} \sigma_x^{(2)} =
-\sigma_z^{(2)}.$ Averaging the EVs obtained immediately after
applying the algorithm unitary operation,
Eq.~(\ref{eq:before_corr21}), and immediately after the correlation
operation, Eq.~(\ref{eq:after_corr21}), gives
\begin{equation}
 \left< \sigma_z^{(2)} \right> = \frac{1}{M} 
                                 \sum_{x \in S, x_1=1}(-1)^{x_2},
\end{equation}
which is identical to an EV performed on the state $1/\sqrt{M} \sum_{x
\in S, x_1=1} \ket{x}$. Combining the two runs of the algorithm, one
with and the other without the correlation operation applied after the
algorithm unitary operation has effectively filtered out the the
desired portion of the state.

When $\left< \sigma_z^{(1)} \right> \geq 0$ the filtering is
conditional on $s_1=0$. The only modification to the scheme described
above is that the correlation operation is
\begin{equation}
 \hat{C}_2(s_1=0) \ket{x_L \ldots x_1} := \left(\sigma^{(2)}_x \right)^{x_1}
                                          \ket{x_L \ldots x_1}
\end{equation}
or equivalently $\hat{C}_2(s_1=0) = \sigma_x^{(1)} \, \hat{C}_2(s_1=1)
\, \sigma_x^{(1)}$.  The average of the EVs from the two runs will be
identical to that performed on $1/\sqrt{M} \sum_{x \in S, x_1=0}
\ket{x}$.

This first round of filtered EVs will give $s_2$ and $s_1$ for some $s
\in S$. This procedure can be iterated with appropriate modifications
of the correlation operation, eventually yielding the location of a
single marked item.  The general idea is demonstrated by considering
the situation where the filtering is conditional on $s_k = s_{k-1} =
\ldots s_1 = 1$. The relevant correlation operator is defined via
\begin{widetext}
\begin{equation}
 \hat{C}_{k+1}(s_k=1,\ldots,s_1=1) \ket{x_L \ldots x_1} := 
                 \left(\sigma^{(k+1)}_x \right)^{g(x_k\ldots x_1)}
                 \ket{x_L \ldots x_1}
\end{equation}
\end{widetext}
where
\begin{equation}
 g(x_k\ldots x_1) := x_kx_{k-1}\ldots x_2x_1 \oplus 1.
 \label{eq:corr_function}
\end{equation}
Averaging the EVs obtained in the two cases where
$\hat{U}_{\text{alg}}$ and $\hat{C}_{k+1}(s_k=1,\ldots,s_1=1)\,
\hat{U}_{\text{alg}}$ are applied to $\ket{\psi_i}$ yields
\begin{equation}
 \left< \sigma_z^{(k+1)} \right> = \frac{1}{M} 
                                   \sum_{x \in S, x_k = 1,  \ldots , x_1=1}
                                   (-1)^{x_{k+1}}
\end{equation}
and this achieves the desired filtration. Instances where any of the
filtering conditions is of the form $s_j = 0$ are accommodated by
using
%
\begin{equation}
 \hat{C}_{k+1}(\ldots ,s_j=0, \ldots ) = \sigma_x^{(j)} \, 
                 \hat{C}_{k+1}(\ldots ,s_j=1, \ldots ) \, 
                                         \sigma_x^{(j)}.
\end{equation}
%
Note that $g$ defined in Eq.~(\ref{eq:corr_function}) can be computed
with $k$ multiplications and a single additions, which implies that
the filtering procedure requires $O(\log_2{N})$ additional operations
at worst.

This filtered EV technique allows an expectation value quantum
computer to use Grover's algorithm to locate at least one marked item
with certainty. The additional cost is that the algorithm must be run
repeatedly and the relevant correlation operations must be
included. However, the number of runs of the algorithm is not
prohibitive since at each level of filtration only one additional run
is required (to compare with the results from the previous
level). Therefore at most $\log_2{N}$ runs of the algorithm are needed
to locate a single marked item with certainty.

This EV version of Grover's algorithm is also amenable to the
treatment which resulted in the truncated EV version of the algorithm
with only one marked item. Again suppose that $\hat{U}_{\text{alg}}$
is modified so that it terminates after $m$ applications of
$\hat{G}$. Applying this to the system in the state $\ket{\psi_i}=
1/\sqrt{2^L}\sum_{x=0}^{2L-1}\ket{x}$ and using the filtering
procedure to select a subset $S' \subset S$ is easily shown to result
in
\begin{equation}
 \left< \sigma_z^{(k)} \right>  =  \frac{A_m}{M} \sum_{x \in S'} (-1)^{x_k}
 \label{eq:trunc_grover_filtered}
\end{equation}
where the EV attenuation is
\begin{equation}
  A_m = \frac{\sin^2{[(2m+1)\theta/2]}\; N - M}{N-M}.
 \label{eq:Mamp_after_m}
\end{equation}
The only effect of truncating $\hat{U}_{\text{alg}}$ is a reduction in
the magnitudes of the EVs by a factor of $A_m$ and all the filtered EV
procedures described in this section for locating a marked item still
apply. Thus the truncated EV version of the algorithm applies for
searching a database containing any number of marked items and
whenever the basic measurement quantities are expectation values it is
superior to the standard (with filtering as necessary) EV version.

The minimum number of applications of $\hat{G}$ must be such that it
yields sufficiently large EVs whenever the filtering scheme leaves
only one marked item. Here it is essential to be able to distinguish
between $\left< \sigma_z^{(k)} \right> = +A_m/M$ and $\left<
\sigma_z^{(k)} \right> = -A_m/M$. If this is true then it will be
possible to discern bit values for marked items reliably even when the
filtered EV approach leaves more than one marked item. In the standard
EV version $A_m \approx 1$ and in order to locate a marked item with
certainty the maximum tolerable error is $\epsilon_{\text{stand}} :=
1/M$. Now suppose that for a particular realization of the computing
device, expectation values are determined to an accuracy of $\epsilon
< \epsilon_{\text{stand}}$. Then it will be possible to successfully
infer the location of a marked item after $m$ applications of
$\hat{G}$ provided that $A_m/M >\epsilon.$
Equation~(\ref{eq:Mamp_after_m}) implies that this is true if
\begin{equation}
 m > \frac{1}{\theta} \arcsin{\sqrt{\epsilon/\epsilon_{\text{stand}} 
                           + (1-\epsilon/\epsilon_{\text{stand}})M/N}} 
                           - \frac{1}{2}.
\end{equation}
If $M \ll N$ then $\theta \approx 2\sqrt{M/N}$ and the minimum number
of applications of $\hat{G}$ for the truncated EV version is
\begin{equation}
 m_{\text{trunc}} \approx \frac{1}{2} 
                          \sqrt{\frac{N}{M}} 
                          \arcsin{\sqrt{\epsilon/\epsilon_{\text{stand}} 
                          + (1-\epsilon/\epsilon_{\text{stand}})M/N}}.
\end{equation}
The number of applications of $\hat{G}$ required for the standard EV
version to locate a marked item with certainty is $m_{\text{stand}}
\approx \pi/4 \sqrt{N/M}$ and thus
\begin{equation}
 m_{\text{trunc}} \approx  m_{\text{stand}} \frac{2}{\pi} 
                         \arcsin{\sqrt{\epsilon/\epsilon_{\text{stand}} 
                        + (1-\epsilon/\epsilon_{\text{stand}})M/N}}.
 \label{eq:final_ratio}
\end{equation}
Clearly for a given number of marked items,
$\epsilon/\epsilon_{\text{stand}}$ is the determining factor in the
reduction in the minimum number of applications of $\hat{G}$, which
increases monotonically, with respect to this parameter, from $2
m_{\text{stand}} \arcsin{\sqrt{M/N}} / \pi$ (for
$\epsilon/\epsilon_{\text{stand}} = 0$) to $m_{\text{stand}}$ (for
$\epsilon/\epsilon_{\text{stand}} = 1$). If $\epsilon \ll
\epsilon_{\text{stand}}$ and $M \ll N$ then
\begin{equation}
 \frac{m_{\text{trunc}}}{m_{\text{stand}}} \approx 
              \frac{2}{\pi}\sqrt{\epsilon/\epsilon_{\text{stand}} + M/N}.
\end{equation} 

One remaining issue is the relationship between the ensemble's size
and it's ability to estimate expectation values
accurately. Determining expectation values requires an ensemble of
identically prepared quantum systems, to each member of which the same
experiment (i.e.\ the same sequence of unitary transformations and
projective measurements) is applied. In this context, expectation
values are identified with sample averages calculated from outcomes of
projective measurements, one per member of the ensemble. In particular
if the experiment is performed on an ensemble consisting of $n$
members and the number of times that the PMs on qubit $k$ return $0$
and $1$ are $n_0$ and $n_1$ respectively then the expectation value
$\left< \sigma_z^{(k)}\right>$ is equated with the sample average
$\overline{x}_k \equiv (n_0 - n_1)/n$. Here the distribution is
binomial, in which case Chebyshev's inequality \cite{feller} offers
\begin{equation}
 \text{Prob} \left\{  \left| \overline{x}_k - \left< \sigma_z^{(k)}\right> 
                    \right| < \epsilon\right\} > 1- 1/4n\epsilon^2
\end{equation}
where $\epsilon>0$ bounds the error in the estimate. In realizations
such as standard room temperature, solution state NMR, $n \approx
10^{20}$ and estimates within very small error bounds are easily
attained with near certainty. Thus the ensemble size is of little
concern for standard NMR approaches although it may become an issue
elsewhere.

\section{Other algorithms}
\label{sec:other}

It remains to ask whether such modifications apply to other quantum
algorithms. It is unlikely that the truncation will be applicable
elsewhere as it requires an iterated operation, which amongst quantum
algorithms is unique to Grover's algorithm. On the other hand the
filtered EV scheme could perhaps be more widely applicable. However,
initial attempts to apply it to Shor's factorization algorithm have
been unsuccessful. The quantum part of Shor's algorithm \cite{shor}
seeks to find the period, $r$, of a well defined function of the
number to be factorized. For the present discussion the essential
details are as follows. After applying the unitary algorithm operator,
a computational basis measurement is performed on an $L$, suitably
large, qubit data register. The result of this is, with high
probability, an integer $y \approx 2^L k/r$ for any one of
$k=0,1,\ldots , r-1$. All possibilities for $k$ are approximately
equally likely (i.~e.\ $\approx 1/r$). An appropriately terminated
continued fractions approximation to $y/2^L$ gives $k/r$ and if $k$
and $r$ are relatively prime, which occurs sufficiently often, then
the denominator of the continued fractions approximation gives $r$. It
is well known that simply substituting EVs for the PMs will not work
\cite{GC}. One proposed solution is to have the quantum computer
perform the continued fractions expansion reversibly and follow this
with EVs \cite{GC}. Although these additional steps are still
computationally efficient, it would be preferable to avoid them. One
attempt to do so would be to use filtered EVs to isolate the
computational basis states in the region of $y \approx 2^L k/r$ for
some desirable choice of $k$. However, although there is a reasonable
hope that, by using a variant of the binary search, this will succeed
with an sufficiently small number of additional steps this approach is
still futile since the amplitudes of the resulting EVs will scale as
$\sim 1/r$ and hence diminish exponentially as the size of $r$
increases. This straightforward approach appears to be doomed and no
alternatives are clearly visible.

\section{Conclusions}

In conclusion we have described in detail modifications which
translate Grover's search algorithm from a formulation in terms of
single quantum systems amenable to projective measurements into one
suitable for ensemble quantum computers where only expectation values
are available. This included a thorough exposition of a filtered
expectation value technique, which has the potential to be applicable
more generally. In the context of Grover's algorithm, we presented a
significantly modified version of the algorithm which, for
realizations involving expectation values, is superior to the standard
version. The existence of this superior version indicates that it may
be worthwhile to investigate similar modifications to other quantum
algorithms although our initial attempts to do so have failed.

\begin{acknowledgments}
 The author would like to thank Bob Griffiths for many useful discussions. This work was supported by NSF grant PHY 99-00755.
\end{acknowledgments}

\bibliography{paper3}

\begin{thebibliography}{27}
\expandafter\ifx\csname natexlab\endcsname\relax\def\natexlab#1{#1}\fi
\expandafter\ifx\csname bibnamefont\endcsname\relax
  \def\bibnamefont#1{#1}\fi
\expandafter\ifx\csname bibfnamefont\endcsname\relax
  \def\bibfnamefont#1{#1}\fi
\expandafter\ifx\csname citenamefont\endcsname\relax
  \def\citenamefont#1{#1}\fi
\expandafter\ifx\csname url\endcsname\relax
  \def\url#1{\texttt{#1}}\fi
\expandafter\ifx\csname urlprefix\endcsname\relax\def\urlprefix{URL }\fi
\providecommand{\bibinfo}[2]{#2}
\providecommand{\eprint}[2][]{\url{#2}}

\bibitem[{\citenamefont{Cleve et~al.}(1998)\citenamefont{Cleve, Ekert,
  Macchiavello, and Mosca}}]{cleve}
\bibinfo{author}{\bibfnamefont{R.}~\bibnamefont{Cleve}},
  \bibinfo{author}{\bibfnamefont{A.}~\bibnamefont{Ekert}},
  \bibinfo{author}{\bibfnamefont{C.}~\bibnamefont{Macchiavello}},
  \bibnamefont{and} \bibinfo{author}{\bibfnamefont{M.}~\bibnamefont{Mosca}},
  \bibinfo{journal}{Proc.\ R.\ Soc.\ Lond. A} \textbf{\bibinfo{volume}{454}},
  \bibinfo{pages}{339} (\bibinfo{year}{1998}).

\bibitem[{\citenamefont{Shor}(1997)}]{shor}
\bibinfo{author}{\bibfnamefont{P.}~\bibnamefont{Shor}}, \bibinfo{journal}{SIAM
  J.\ Comput.} \textbf{\bibinfo{volume}{26}}, \bibinfo{pages}{1484}
  (\bibinfo{year}{1997}).

\bibitem[{\citenamefont{Grover}(1997)}]{grov}
\bibinfo{author}{\bibfnamefont{L.~K.} \bibnamefont{Grover}},
  \bibinfo{journal}{Phys.\ Rev.\ Lett.} \textbf{\bibinfo{volume}{79}},
  \bibinfo{pages}{325} (\bibinfo{year}{1997}).

\bibitem[{\citenamefont{Cory et~al.}(1997)\citenamefont{Cory, Fahmy, and
  Havel}}]{CFH}
\bibinfo{author}{\bibfnamefont{D.~G.} \bibnamefont{Cory}},
  \bibinfo{author}{\bibfnamefont{A.~F.} \bibnamefont{Fahmy}}, \bibnamefont{and}
  \bibinfo{author}{\bibfnamefont{T.~F.} \bibnamefont{Havel}},
  \bibinfo{journal}{Proc.\ Nat.\ Acad.\ Sci.} \textbf{\bibinfo{volume}{94}},
  \bibinfo{pages}{1634} (\bibinfo{year}{1997}).

\bibitem[{\citenamefont{Chuang et~al.}(1998{\natexlab{a}})\citenamefont{Chuang,
  Gershenfeld, Kubinec, and Leung}}]{ChGKL}
\bibinfo{author}{\bibfnamefont{I.~L.} \bibnamefont{Chuang}},
  \bibinfo{author}{\bibfnamefont{N.}~\bibnamefont{Gershenfeld}},
  \bibinfo{author}{\bibfnamefont{M.~G.} \bibnamefont{Kubinec}},
  \bibnamefont{and} \bibinfo{author}{\bibfnamefont{D.~W.} \bibnamefont{Leung}},
  \bibinfo{journal}{Proc.\ R.\ Soc.\ Lond. A} \textbf{\bibinfo{volume}{454}},
  \bibinfo{pages}{447} (\bibinfo{year}{1998}{\natexlab{a}}).

\bibitem[{\citenamefont{Chuang et~al.}(1998{\natexlab{b}})\citenamefont{Chuang,
  Gershenfeld, and Kubinec}}]{ChGK98}
\bibinfo{author}{\bibfnamefont{I.~L.} \bibnamefont{Chuang}},
  \bibinfo{author}{\bibfnamefont{N.}~\bibnamefont{Gershenfeld}},
  \bibnamefont{and} \bibinfo{author}{\bibfnamefont{M.}~\bibnamefont{Kubinec}},
  \bibinfo{journal}{Phys. Rev. Lett.} \textbf{\bibinfo{volume}{80}},
  \bibinfo{pages}{3408} (\bibinfo{year}{1998}{\natexlab{b}}).

\bibitem[{\citenamefont{Jones et~al.}(1998)\citenamefont{Jones, Mosca, and
  Hansen}}]{JMH98}
\bibinfo{author}{\bibfnamefont{J.~A.} \bibnamefont{Jones}},
  \bibinfo{author}{\bibfnamefont{M.}~\bibnamefont{Mosca}}, \bibnamefont{and}
  \bibinfo{author}{\bibfnamefont{R.~H.} \bibnamefont{Hansen}},
  \bibinfo{journal}{Nature} \textbf{\bibinfo{volume}{399}},
  \bibinfo{pages}{344} (\bibinfo{year}{1998}).

\bibitem[{\citenamefont{Linden et~al.}(1998)\citenamefont{Linden, Barjat, and
  Freeman}}]{linden98}
\bibinfo{author}{\bibfnamefont{N.}~\bibnamefont{Linden}},
  \bibinfo{author}{\bibfnamefont{H.}~\bibnamefont{Barjat}}, \bibnamefont{and}
  \bibinfo{author}{\bibfnamefont{R.}~\bibnamefont{Freeman}},
  \bibinfo{journal}{Chem. Phys. Lett.} \textbf{\bibinfo{volume}{296,}},
  \bibinfo{pages}{61} (\bibinfo{year}{1998}).

\bibitem[{\citenamefont{Vandersypen et~al.}(1999)\citenamefont{Vandersypen,
  Yannoni, Sherwood, and Chuang}}]{VYSCh99}
\bibinfo{author}{\bibfnamefont{L.~M.~K.} \bibnamefont{Vandersypen}},
  \bibinfo{author}{\bibfnamefont{C.~S.} \bibnamefont{Yannoni}},
  \bibinfo{author}{\bibfnamefont{M.~H.} \bibnamefont{Sherwood}},
  \bibnamefont{and} \bibinfo{author}{\bibfnamefont{I.~L.}
  \bibnamefont{Chuang}}, \bibinfo{journal}{Phys. Rev. Lett.}
  \textbf{\bibinfo{volume}{83}}, \bibinfo{pages}{3085} (\bibinfo{year}{1999}).

\bibitem[{\citenamefont{Vandersypen
  et~al.}(2000{\natexlab{a}})\citenamefont{Vandersypen, Steffen, Sherwood,
  Yannoni, Breyta, and Chuang}}]{VSSYBCh00}
\bibinfo{author}{\bibfnamefont{L.~M.~K.} \bibnamefont{Vandersypen}},
  \bibinfo{author}{\bibfnamefont{M.}~\bibnamefont{Steffen}},
  \bibinfo{author}{\bibfnamefont{M.~H.} \bibnamefont{Sherwood}},
  \bibinfo{author}{\bibfnamefont{C.~S.} \bibnamefont{Yannoni}},
  \bibinfo{author}{\bibfnamefont{G.}~\bibnamefont{Breyta}}, \bibnamefont{and}
  \bibinfo{author}{\bibfnamefont{I.~L.} \bibnamefont{Chuang}},
  \bibinfo{journal}{App. Phys. Lett.} \textbf{\bibinfo{volume}{76}},
  \bibinfo{pages}{646} (\bibinfo{year}{2000}{\natexlab{a}}).

\bibitem[{\citenamefont{Marx et~al.}(2000)\citenamefont{Marx, Fahmy, Myers,
  Bermel, and Glaser}}]{marx00}
\bibinfo{author}{\bibfnamefont{R.}~\bibnamefont{Marx}},
  \bibinfo{author}{\bibfnamefont{A.~F.} \bibnamefont{Fahmy}},
  \bibinfo{author}{\bibfnamefont{J.~M.} \bibnamefont{Myers}},
  \bibinfo{author}{\bibfnamefont{W.}~\bibnamefont{Bermel}}, \bibnamefont{and}
  \bibinfo{author}{\bibfnamefont{S.~J.} \bibnamefont{Glaser}},
  \bibinfo{journal}{Phys. Rev. A} \textbf{\bibinfo{volume}{62}},
  \bibinfo{pages}{012310} (\bibinfo{year}{2000}).

\bibitem[{\citenamefont{Cummins and Jones}(2000)}]{cummins00}
\bibinfo{author}{\bibfnamefont{H.~K.} \bibnamefont{Cummins}} \bibnamefont{and}
  \bibinfo{author}{\bibfnamefont{J.~A.} \bibnamefont{Jones}},
  \bibinfo{journal}{New J. Phys.} \textbf{\bibinfo{volume}{2}},
  \bibinfo{pages}{6} (\bibinfo{year}{2000}).

\bibitem[{\citenamefont{Collins et~al.}(2000)\citenamefont{Collins, Kim,
  Holton, Sierzputowska-Grazc, and Stejskal}}]{collins00}
\bibinfo{author}{\bibfnamefont{D.}~\bibnamefont{Collins}},
  \bibinfo{author}{\bibfnamefont{K.~W.} \bibnamefont{Kim}},
  \bibinfo{author}{\bibfnamefont{W.~C.} \bibnamefont{Holton}},
  \bibinfo{author}{\bibfnamefont{H.}~\bibnamefont{Sierzputowska-Grazc}},
  \bibnamefont{and} \bibinfo{author}{\bibfnamefont{E.~O.}
  \bibnamefont{Stejskal}}, \bibinfo{journal}{Phys. Rev. A}
  \textbf{\bibinfo{volume}{62}}, \bibinfo{pages}{022304}
  (\bibinfo{year}{2000}).

\bibitem[{\citenamefont{Kim et~al.}(2000)\citenamefont{Kim, Lee, and
  Lee}}]{kim00}
\bibinfo{author}{\bibfnamefont{J.}~\bibnamefont{Kim}},
  \bibinfo{author}{\bibfnamefont{J.-S.} \bibnamefont{Lee}}, \bibnamefont{and}
  \bibinfo{author}{\bibfnamefont{S.}~\bibnamefont{Lee}},
  \bibinfo{journal}{Phys. Rev. A} \textbf{\bibinfo{volume}{62}},
  \bibinfo{pages}{022312} (\bibinfo{year}{2000}).

\bibitem[{\citenamefont{Vandersypen
  et~al.}(2000{\natexlab{b}})\citenamefont{Vandersypen, Steffen, Breyta,
  Yannoni, Cleve, and Chuang}}]{vandersypen00}
\bibinfo{author}{\bibfnamefont{L.~M.~K.} \bibnamefont{Vandersypen}},
  \bibinfo{author}{\bibfnamefont{M.}~\bibnamefont{Steffen}},
  \bibinfo{author}{\bibfnamefont{G.}~\bibnamefont{Breyta}},
  \bibinfo{author}{\bibfnamefont{C.~S.} \bibnamefont{Yannoni}},
  \bibinfo{author}{\bibfnamefont{R.}~\bibnamefont{Cleve}}, \bibnamefont{and}
  \bibinfo{author}{\bibfnamefont{I.~L.} \bibnamefont{Chuang}},
  \bibinfo{journal}{Phys. Rev. Lett.} \textbf{\bibinfo{volume}{85}},
  \bibinfo{pages}{5452} (\bibinfo{year}{2000}{\natexlab{b}}).

\bibitem[{\citenamefont{Weinstein et~al.}(2001)\citenamefont{Weinstein, Pravia,
  Fortunato, Lloyd, and Cory}}]{weinstein01}
\bibinfo{author}{\bibfnamefont{Y.~S.} \bibnamefont{Weinstein}},
  \bibinfo{author}{\bibfnamefont{M.~A.} \bibnamefont{Pravia}},
  \bibinfo{author}{\bibfnamefont{E.~M.} \bibnamefont{Fortunato}},
  \bibinfo{author}{\bibfnamefont{S.}~\bibnamefont{Lloyd}}, \bibnamefont{and}
  \bibinfo{author}{\bibfnamefont{D.~G.} \bibnamefont{Cory}},
  \bibinfo{journal}{Phys. Rev. Lett.} \textbf{\bibinfo{volume}{86}},
  \bibinfo{pages}{1889} (\bibinfo{year}{2001}).

\bibitem[{\citenamefont{Long et~al.}(2001)\citenamefont{Long, Yan, Li, Tu, Tao,
  Chen, Liu, Zhang, Luo, Xiao et~al.}}]{long01}
\bibinfo{author}{\bibfnamefont{G.~L.} \bibnamefont{Long}},
  \bibinfo{author}{\bibfnamefont{H.~Y.} \bibnamefont{Yan}},
  \bibinfo{author}{\bibfnamefont{Y.~S.} \bibnamefont{Li}},
  \bibinfo{author}{\bibfnamefont{C.~C.} \bibnamefont{Tu}},
  \bibinfo{author}{\bibfnamefont{J.~X.} \bibnamefont{Tao}},
  \bibinfo{author}{\bibfnamefont{H.~M.} \bibnamefont{Chen}},
  \bibinfo{author}{\bibfnamefont{M.~L.} \bibnamefont{Liu}},
  \bibinfo{author}{\bibfnamefont{X.}~\bibnamefont{Zhang}},
  \bibinfo{author}{\bibfnamefont{J.}~\bibnamefont{Luo}},
  \bibinfo{author}{\bibfnamefont{L.}~\bibnamefont{Xiao}}, \bibnamefont{et~al.},
  \bibinfo{journal}{Phys. Lett. A} \textbf{\bibinfo{volume}{286}},
  \bibinfo{pages}{121} (\bibinfo{year}{2001}).

\bibitem[{\citenamefont{Gershenfeld and Chuang}(1997)}]{GC}
\bibinfo{author}{\bibfnamefont{N.~A.} \bibnamefont{Gershenfeld}}
  \bibnamefont{and} \bibinfo{author}{\bibfnamefont{I.~L.}
  \bibnamefont{Chuang}}, \bibinfo{journal}{Science}
  \textbf{\bibinfo{volume}{275}}, \bibinfo{pages}{350} (\bibinfo{year}{1997}).

\bibitem[{\citenamefont{Knill et~al.}(1997)\citenamefont{Knill, Chuang, and
  Laflamme}}]{KChL97}
\bibinfo{author}{\bibfnamefont{E.}~\bibnamefont{Knill}},
  \bibinfo{author}{\bibfnamefont{I.}~\bibnamefont{Chuang}}, \bibnamefont{and}
  \bibinfo{author}{\bibfnamefont{R.}~\bibnamefont{Laflamme}},
  \bibinfo{journal}{Phys. Rev. A} \textbf{\bibinfo{volume}{57}},
  \bibinfo{pages}{3348} (\bibinfo{year}{1997}).

\bibitem[{\citenamefont{Cory et~al.}(1998)\citenamefont{Cory, Price, and
  Havel}}]{cory98}
\bibinfo{author}{\bibfnamefont{D.~G.} \bibnamefont{Cory}},
  \bibinfo{author}{\bibfnamefont{M.~D.} \bibnamefont{Price}}, \bibnamefont{and}
  \bibinfo{author}{\bibfnamefont{T.~F.} \bibnamefont{Havel}},
  \bibinfo{journal}{Physica D} \textbf{\bibinfo{volume}{120}},
  \bibinfo{pages}{82} (\bibinfo{year}{1998}).

\bibitem[{\citenamefont{Schack and Caves}(1999)}]{schack99}
\bibinfo{author}{\bibfnamefont{R.}~\bibnamefont{Schack}} \bibnamefont{and}
  \bibinfo{author}{\bibfnamefont{C.~M.} \bibnamefont{Caves}},
  \bibinfo{journal}{Phys. Rev. A} \textbf{\bibinfo{volume}{60}},
  \bibinfo{pages}{4354} (\bibinfo{year}{1999}).

\bibitem[{\citenamefont{Nielsen and Chuang}(2000)}]{NC}
\bibinfo{author}{\bibfnamefont{M.~A.} \bibnamefont{Nielsen}} \bibnamefont{and}
  \bibinfo{author}{\bibfnamefont{I.~L.} \bibnamefont{Chuang}},
  \emph{\bibinfo{title}{Quantum Computation and Quantum Information}}
  (\bibinfo{publisher}{Cambridge University Press},
  \bibinfo{address}{Cambridge}, \bibinfo{year}{2000}).

\bibitem[{\citenamefont{Brassard et~al.}(1998)\citenamefont{Brassard, H{\o}yer,
  and Tapp}}]{brassard98}
\bibinfo{author}{\bibfnamefont{G.}~\bibnamefont{Brassard}},
  \bibinfo{author}{\bibfnamefont{P.}~\bibnamefont{H{\o}yer}}, \bibnamefont{and}
  \bibinfo{author}{\bibfnamefont{A.}~\bibnamefont{Tapp}},
  \bibinfo{journal}{Lect. Notes Comp. Sci.} \textbf{\bibinfo{volume}{1443}},
  \bibinfo{pages}{830} (\bibinfo{year}{1998}).

\bibitem[{\citenamefont{Boyer et~al.}(1998)\citenamefont{Boyer, Brassard,
  H{\o}yer, and Tapp}}]{bbht98}
\bibinfo{author}{\bibfnamefont{M.}~\bibnamefont{Boyer}},
  \bibinfo{author}{\bibfnamefont{G.}~\bibnamefont{Brassard}},
  \bibinfo{author}{\bibfnamefont{P.}~\bibnamefont{H{\o}yer}}, \bibnamefont{and}
  \bibinfo{author}{\bibfnamefont{A.}~\bibnamefont{Tapp}},
  \bibinfo{journal}{Fortschr. Phys.} \textbf{\bibinfo{volume}{46}},
  \bibinfo{pages}{493} (\bibinfo{year}{1998}).

\bibitem[{\citenamefont{Zalka}(1999)}]{zalka99}
\bibinfo{author}{\bibfnamefont{C.}~\bibnamefont{Zalka}},
  \bibinfo{journal}{Phys. Rev. A} \textbf{\bibinfo{volume}{60}},
  \bibinfo{pages}{2746} (\bibinfo{year}{1999}).

\bibitem[{\citenamefont{Bennett et~al.}(1997)\citenamefont{Bennett, Bernstein,
  Brassard, and Vazirani}}]{bbbv97}
\bibinfo{author}{\bibfnamefont{C.~H.} \bibnamefont{Bennett}},
  \bibinfo{author}{\bibfnamefont{E.}~\bibnamefont{Bernstein}},
  \bibinfo{author}{\bibfnamefont{G.}~\bibnamefont{Brassard}}, \bibnamefont{and}
  \bibinfo{author}{\bibfnamefont{U.}~\bibnamefont{Vazirani}},
  \bibinfo{journal}{SIAM J. Comput.} \textbf{\bibinfo{volume}{26}},
  \bibinfo{pages}{1510} (\bibinfo{year}{1997}).

\bibitem[{\citenamefont{Feller}(1968)}]{feller}
\bibinfo{author}{\bibfnamefont{W.}~\bibnamefont{Feller}},
  \emph{\bibinfo{title}{An Introduction to Probability Theory and Its
  Applications}} (\bibinfo{publisher}{Wiley}, \bibinfo{address}{New York},
  \bibinfo{year}{1968}), \bibinfo{edition}{3rd} ed.

\end{thebibliography}

\newpage

\end{document}